\documentclass[9pt,twocolumn,twoside]{osajnl}
\usepackage{amsmath,amssymb,graphicx,bm,epstopdf}
\journal{ol} % Choose journal (ao, aop, josaa, josab, ol)

\setboolean{shortarticle}{true} 
\title{Photonic potential for TM waves}

\author[1,*]{Alessandro Alberucci}
\author[2]{Chandroth P. Jisha}
\author[1,3]{Stefan Nolte}

\affil[1]{Institute of Applied Physics, Abbe Center of Photonics, Friedrich-Schiller-Universität Jena, Albert-Einstein-Str. 15, 07745 Jena, Germany }
\affil[2]{Centro de F\'{\i}sica do Porto, Faculdade de Ci\^encias, Universidade do Porto, %R. Campo Alegre 687, 
Porto 4169-007, Portugal}
\affil[3]{Fraunhofer Institute for Applied Optics and Precision Engineering, Albert-Einstein-Straße 7, 07745 Jena, Germany}

\affil[*]{Corresponding author: alessandro.alberucci@gmail.com}

\ociscodes{(260.2110) Electromagnetic optics; (130.5440) Polarization-selective devices; (130.2790) Guided waves.}

\begin{abstract}
We discuss the effective photonic potential for TM waves in inhomogeneous isotropic media. The model provides an easy and intuitive comprehension of form birefringence, paving the way for a new approach on the design of graded-index optical waveguides on nanometric scales. We investigate the application to nanophotonic devices, including integrated nanoscale wave plates and slot waveguides. %Finally, the relationship with the Pancharatnam-Berry phase is found and discussed.
\end{abstract}

\setboolean{displaycopyright}{true}

\begin{document}

\maketitle

In several situations light can be described as a scalar field, including e.g. the propagation of wider-than-wavelength beams in isotropic materials. Under this condition, a  Schr\"odinger equation can be used in lieu of the Maxwell's equations \cite{Lax:1975}. Nonetheless, the scalar approximation fails even in the presence of discontinuous interfaces between two different isotropic materials, causing for example form birefringence in step-index waveguides \cite{Yariv:1997}. 
Currently, there is a continuous effort towards the miniaturization of sub-wavelength optical waveguides \cite{Xu:2004,Barrelet:2004,Guo:2014,Fang:2015,Zhuang:2016,Alcaraz:2018,Maltese:2018}. In fact, one of the main aims of nanophotonics is to shrink optical waveguides as  much as possible \cite{Bogaerts:2005}. But, in sub-wavelength structures, the vectorial nature of light becomes preponderant \cite{Halir:2015}, yielding to a substantial amount of form birefringence even in GRIN (gradient-index) structures \cite{Love:1979,Hu:2007,Gabrielli:2012,Laba:2014}.\\
\indent Here we use the photonic potential for TM waves, recently introduced by Pick and Moiseyev \cite{Pick:2018}, to model the form birefringence in inhomogeneous sub-wavelength structures using a standard Schr\"odinger equation. We first investigate light propagation in GRIN bell-shaped nano-waveguides and their application as ultra-compact polarization rotator \cite{Yang:2008,Corrielli:2014} in integrated optics. We then apply the model to slot waveguides \cite{Almeida:2004} in GRIN geometries. We verify the validity of our results by direct comparison with finite-difference time domain (FDTD) simulations, carried out using the open source simulator MEEP \cite{Oskooi2010}. \\%Finally, we design the equivalent of a GRIN multi-slot waveguide supporting only the TE mode, thus effectively acting like an integrated polarizer \cite{Yang:2008}. \\ 
%In this Letter we introduce and use an effective photonic potential capable to describe waveplate  . \\
\indent We consider an isotropic non-magnetic material ($\mu=1$) in the harmonic regime (e.m. field proportional to $e^{i\omega t}$), but featuring a dielectric constant $\epsilon=\epsilon_0 n^2(x)$ dependent on $x$. It is well known that Maxwell's equations in two dimensions read \cite{Yariv:1997}
\begin{align}
   & \partial_z^2 E_y + \partial_x^2 E_y + k_0^2 n^2(x) E_y=0, \label{eq:TE} \\
   & \partial_z^2 H_y + \partial_x^2 H_y + k_0^2 n^2(x) H_y - \partial_x \log{\epsilon}\  \partial_x H_y=0, 
    \label{eq:TM}
\end{align}
where $k_0=\omega/c$ is the vacuum wavenumber. According to Eqs.~(\ref{eq:TE}-\ref{eq:TM}), the electromagnetic field is always polarization-dependent in an inhomogeneous material (even in the simplest case of a slab waveguide), that is, form birefringence is an intrinsic property of the system. In modern terminology, optical waves are subject to an intrinsic spin-orbit interaction \cite{Bliokh2015} proportional to the geometry-dependent term $\partial_x \log{\epsilon}$, the latter becoming relevant when the refractive index $n(x)$ varies appreciably on distances comparable with the wavelength $\lambda$.\\
\indent To enlighten the formal analogy with respect to the Schr\"odinger equation, we write \eqref{eq:TE} in the form $\partial_z^2 E_y=-\partial_x^2 E_y - V_{TE}(x) E_y$, with $V_{TE}=k_0^2 n^2(x)$. In the paraxial limit, Eq.~(\ref{eq:TM}) closely recalls the Schr\"odinger equation (setting the equivalent Hamiltonian $\hat{H}_{TM}=(\bm{\hat{p}}-\bm{\hat{A}})^2+\hat{V}_{TM}$, with $\bm{\hat{p}}=-i\hat{x}\partial_x$) for a massive particle subject to a scalar ($V$) and a vector potential ($\bm{A}$). Indeed, \eqref{eq:TM} can be recast as 
\begin{align} \label{eq:TM_vector_potential}
   &\partial_z^2 H_y = - \left(\partial_x -\frac{1}{2}\partial_x \log{\epsilon} \right)^2 H_y - \nonumber \\ & \left\{ k_0^2 n^2(x) - \frac{1}{2}\left[  \frac{1}{2} \left( \partial_x \log{\epsilon} \right)^2 -\partial^2_x \log{\epsilon} \right] \right \} H_y.
\end{align}
Hence, the effective vector potential is  %$\bm{\hat{A}}=\frac{i}{2} \partial_x \log{\epsilon} \hat{x}$
 $\bm{\hat{A}}=-i \hat{x} \partial_x \log{n}$ \cite{Longhi:2015} and the effective photonic potential is given by \cite{Pick:2018}
\begin{equation}  \label{eq:potential_TM}
  V_{TM}=V_{TE}+\frac{\partial_x^2 n}{n} - 2 \left(\frac{\partial_x n}{n} \right)^2.
\end{equation}
Finally, the effective vector potential vanishes if the Weyl-like gauge transformation $\psi=H_ye^{-i \int{Adx}}=H_y e^{-\frac{1}{2}\log{\left(\epsilon/\epsilon_0\right)}}=\frac{H_y}{n(x)}$ is applied \cite{Love:1979,Laba:2014}. In fact, the $z$-component of the Poynting vector for the TM wave is $\bm{S}=-\frac{H_y^* \partial_z H_y}{2\omega \epsilon(x)}$, thus the derived scalar field $\psi$ conserves the integral $\mathcal{P}=\int{|\psi|^2dx}$, a property fulfilled by any solution of the Schr\"odinger equation. 
\begin{figure}[t]
\centering
\fbox{\includegraphics[width=\linewidth]{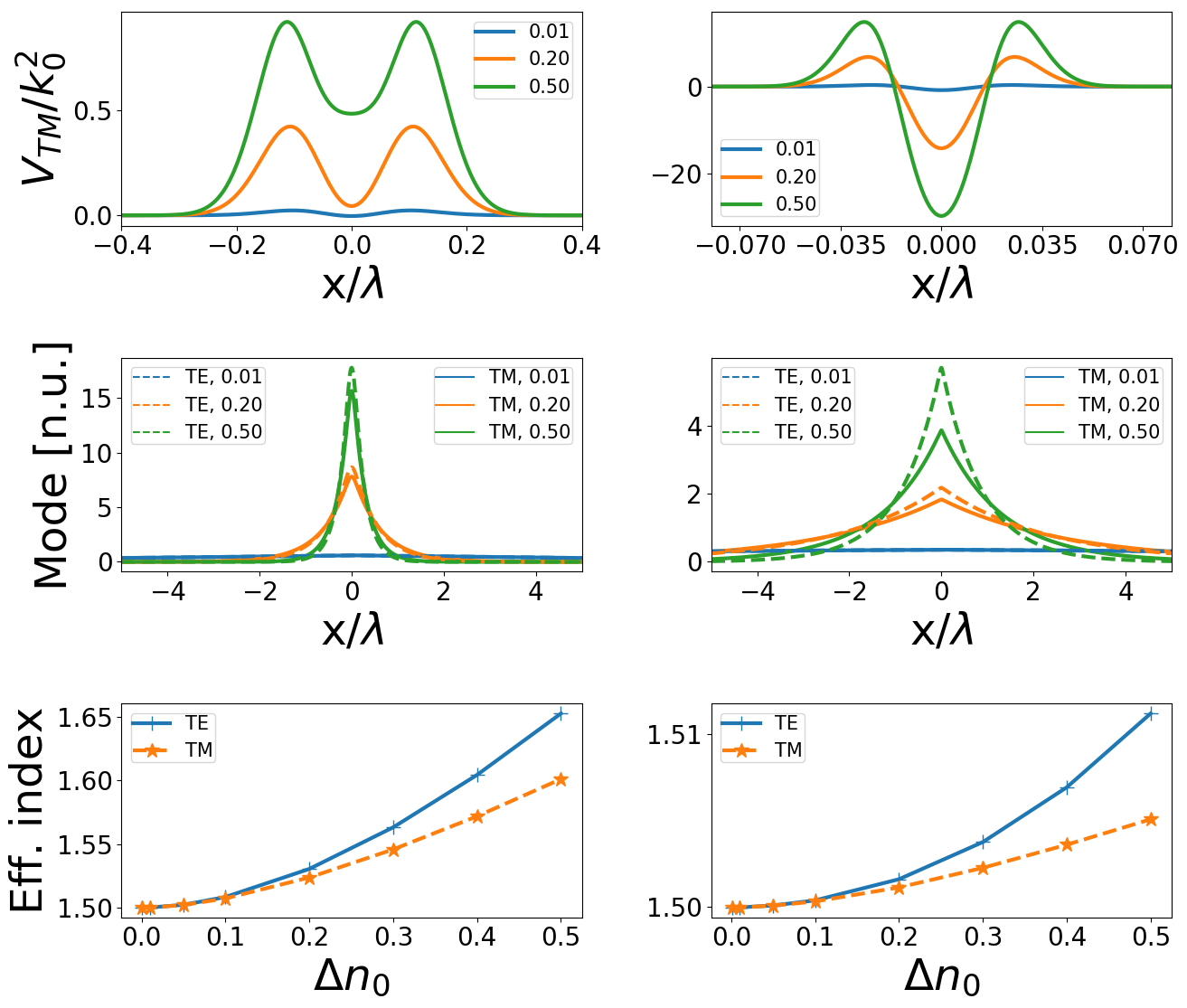}}
\caption{Gaussian nanoguide. Top row: normalized photonic potential $V_{TM}/k_0^2-n_0^2$ versus $x/\lambda$ for three values of $\Delta n_0$ reported in the legend. Middle row: mode profile versus $x/\lambda$ for the TM (solid lines, field $H_y$) and for the TE (dashed lines, $E_y$ component) case, for three values of $\Delta n_0$ reported in the legend. Bottom row: Effective index versus the index well peak $\Delta n_0$ for TE (solid blue line with plus) and TM (dashed orange line with stars) polarization. Left and right column correspond to $w/\lambda=0.1~$ and $w/\lambda=0.02~$, respectively. All the results are computed for $n_0=1.5$.}
\label{fig:bell_shaped}
\end{figure}
\begin{figure}[t]
\centering
\fbox{\includegraphics[width=\linewidth]{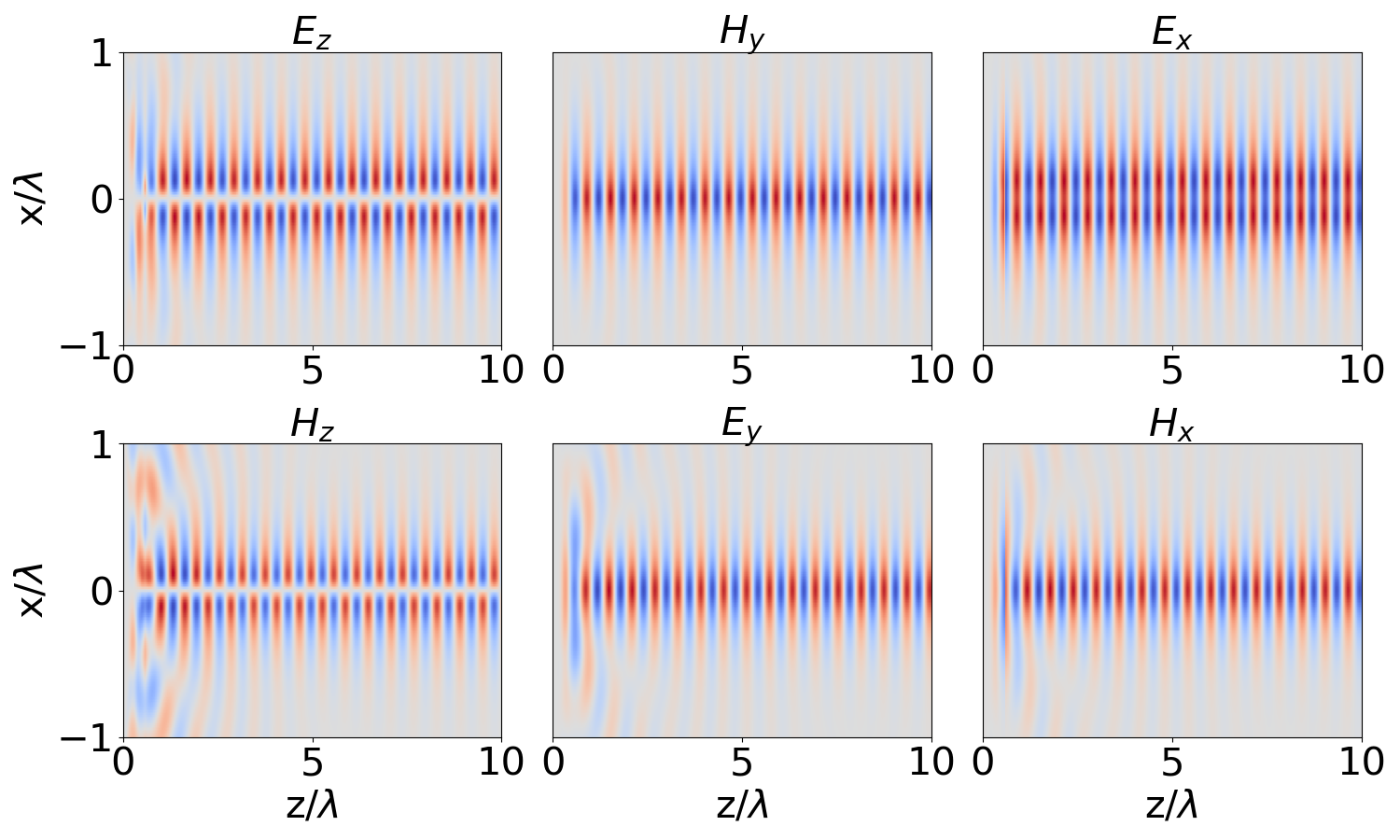}}
\caption{FDTD simulations of a Gaussian nanoguide with $\Delta n_0 = 0.5 $ and $w/\lambda=0.1~$. Top and bottom row show the e.m. field for TM and TE polarization, respectively. Only the fundamental mode is excited into the waveguide, the latter maintaining its profile up to the end of the simulation grid, equal to $50 \lambda$.}
\label{fig:FDTDGauss}
\end{figure} \\
\indent Let us start by considering a bell-shaped index well, for example a Gaussian shape $n=n_0+\Delta n_0 e^{-{x^2}/{w^2}}$. The top row in Fig.~\ref{fig:bell_shaped} \cite{Hunter:2007} shows the potential for the TM polarization. Up to $w\approx \lambda$, $V_{TM}$ is very similar to $V_{TE}$ and the form birefringence is negligible. For $w/\lambda=0.1$, a dip in the center of the effective potential appears, for any value of $\Delta n_0$. Despite that, the value of $V_{TM}$ in $x=0$ stays positive. For $w/\lambda=0.02$ the potential $V_{TM}$ in the center becomes negative. Consequently, the TM mode, plotted in the middle row, becomes wider than for the TE polarization, yielding an increase in the corresponding form birefringence (the corresponding effective indices $n_{TE}$ and $n_{TM}$ are plotted in the bottom row in Fig.~\ref{fig:bell_shaped}). Finally, we note that the optical modes for $w/\lambda=0.02$ are broader than for $w/\lambda=0.1$ due to the diffraction limit \cite{Tong:2004}, corresponding to a lower effective refractive index for the narrowest nanoguide.\\
\indent An example of light propagation in a Gaussian nanoguide is plotted in Fig.~\ref{fig:FDTDGauss}. The TE polarization features single-peak  transverse fields ($E_y$ and $H_x=-\frac{i\partial_z E_y}{\omega \mu_0}$). In the TM case the magnetic field $H_y$ is also single humped, in agreement with Fig.~\ref{fig:bell_shaped}. The electric field $E_x=\frac{i\partial_z H_y}{\omega \epsilon_0 n^2(x)}$ is instead double peaked owing to the fast spatial variations in $n(x)$. Finally, the FDTD profiles match very well with the eigenfunctions computed from \eqref{eq:TE} and \eqref{eq:TM_vector_potential}. With respect to slab waveguides, GRIN structures offer an additional degree of freedom, paving the way to the simultaneous maximization of the form birefringence and of the overlap between TE and TM modes.
\begin{figure}[!bhp]
\centering
\fbox{\includegraphics[width=\linewidth]{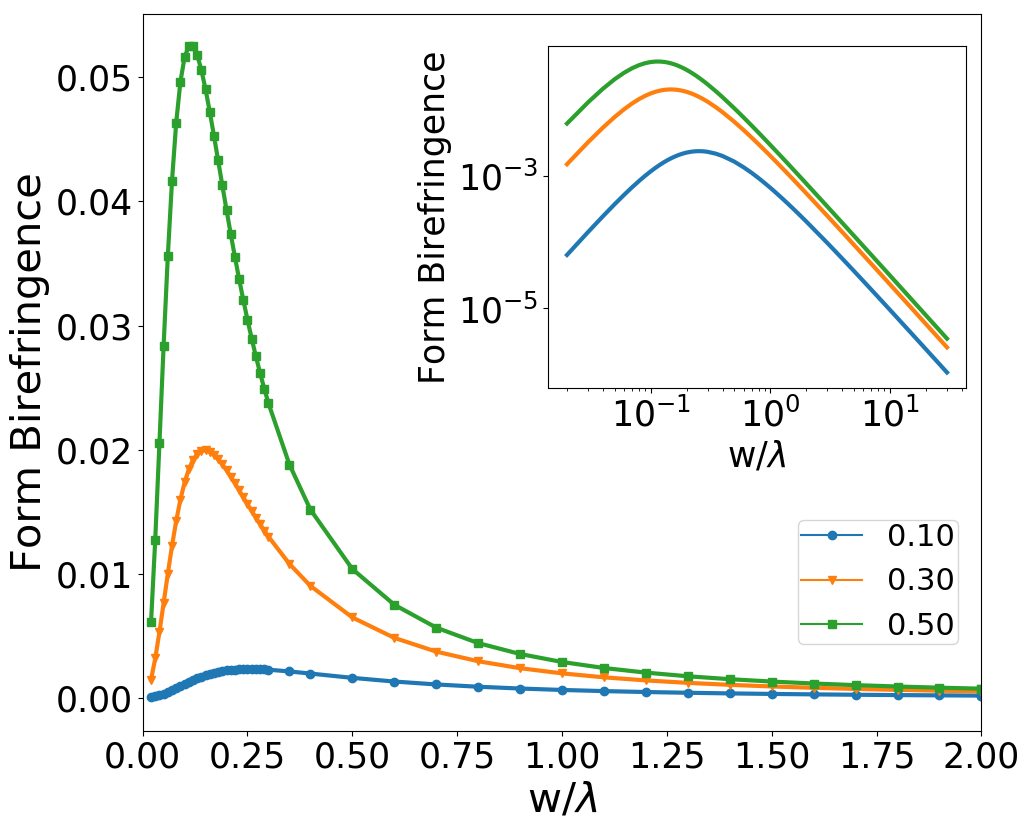}}
\caption{Gaussian nanoguide. Form birefringence $n_{TE}-n_{TM}$ versus the normalized waveguide width $w/\lambda$. In the inset the same curves are replotted in log-log scale. In the legend the corresponding $\Delta n_0$ are reported. All the results are computed for $n_0=1.5$.}
\label{fig:form_birefringence}
\end{figure}
To address the performance of the nanoguide as an integrated waveplate \cite{Yang:2008,Velasco:2012}, in Fig.~\ref{fig:form_birefringence} the form birefringence $n_{TE}-n_{TM}$ is plotted versus the size of the waveguide, both in linear (main panel) and log-log scale (inset). The position of the maximum form birefringence depends on $\Delta n_0$, ranging from $w/\lambda\approx 0.11$ for $\Delta n_0=0.5$, to $w/\lambda\approx 0.25$ for $\Delta n_0=0.1$. For $\Delta n_0=0.5$, the birefringence is 0.05, or $0.1\Delta n_0$. For wide guides, the form birefringence goes as $\propto (w/\lambda)^{-\beta}$ with $\beta=1.95\pm 0.05$, the second decimal digit being dependent on $\Delta n_0$ \cite{Snyder:1986,Corrielli:2014}. For example, these results find applications in the investigation of fs-written waveguides \cite{Corrielli:2014}.
\begin{figure}[htbp]
\centering
\fbox{\includegraphics[width=\linewidth]{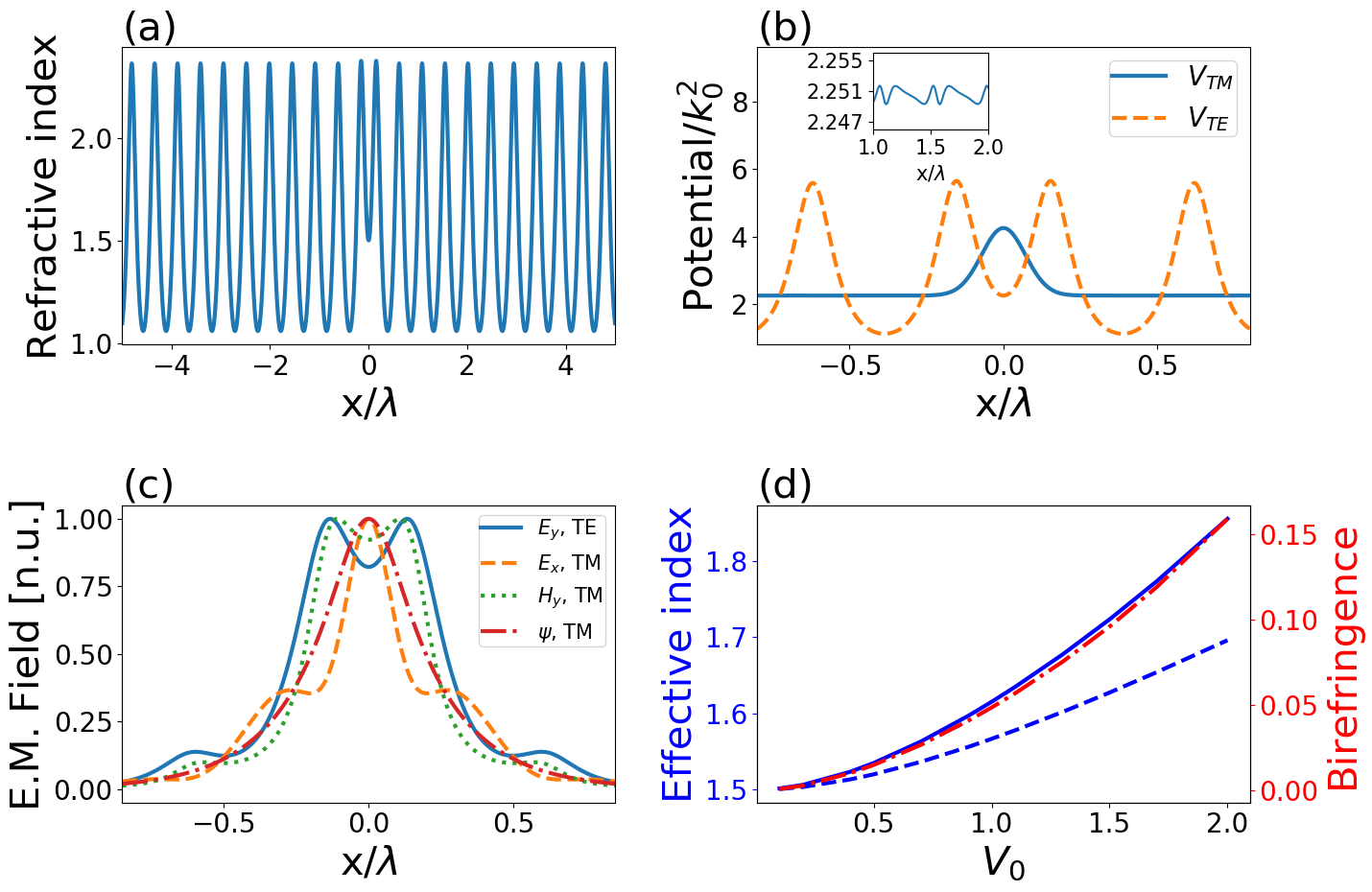}}
\caption{Gaussian potential for the TM polarization. (a) Refractive index and (b) the corresponding photonic potentials versus $x$ for $w/\lambda=0.1$ and $V_0=2$. The inset in (b) provides a magnification of $V_{TM}$ away from the guide. (c) Guided modes for the potentials plotted in (b). (d)  Effective refractive index (left axis, blue color) for the TE (solid line) and the TM polarization (dashed line) versus $V_0$ for $w/\lambda=0.1$; the right axis quantifies the corresponding form birefringence (red dashed-dotted line). We set $n_0=1.5$.}
\label{fig:designed_VTM}
\end{figure}
\begin{figure}[t]
\centering
\fbox{\includegraphics[width=7cm]{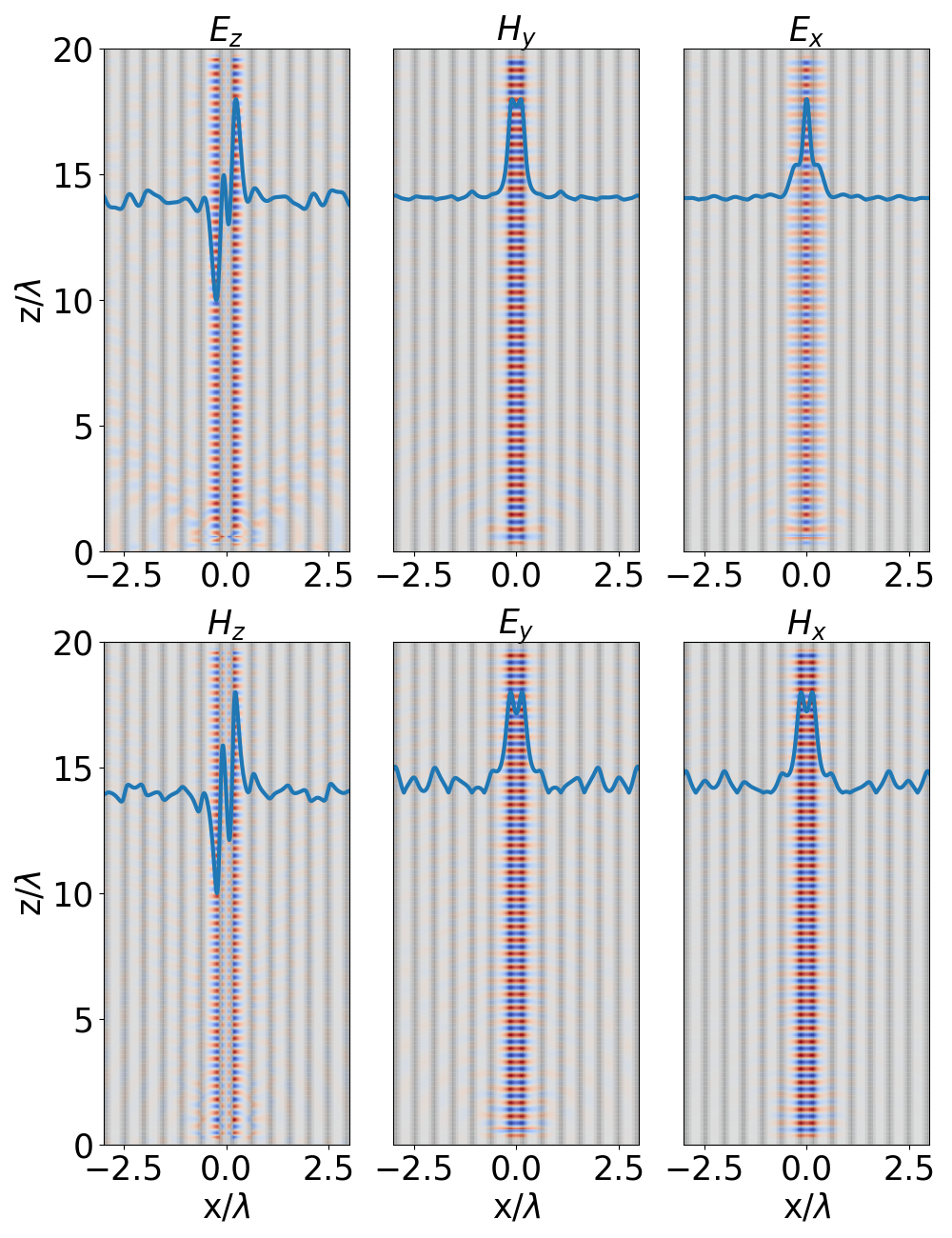}}
\caption{Gaussian potential for the TM polarization. FDTD simulations for the designed waveguide with $V_0=2$. The refractive index modulation is shown in the background. Top row depicts the TM component and bottom row depicts the TE component. The solid curve in each panel is the field amplitude at that position. We fixed $n_0=1.5$.}
\label{fig:designed_VTM_fdtd}
\end{figure}
\begin{figure}[t]
\centering
\fbox{\includegraphics[width=\linewidth]{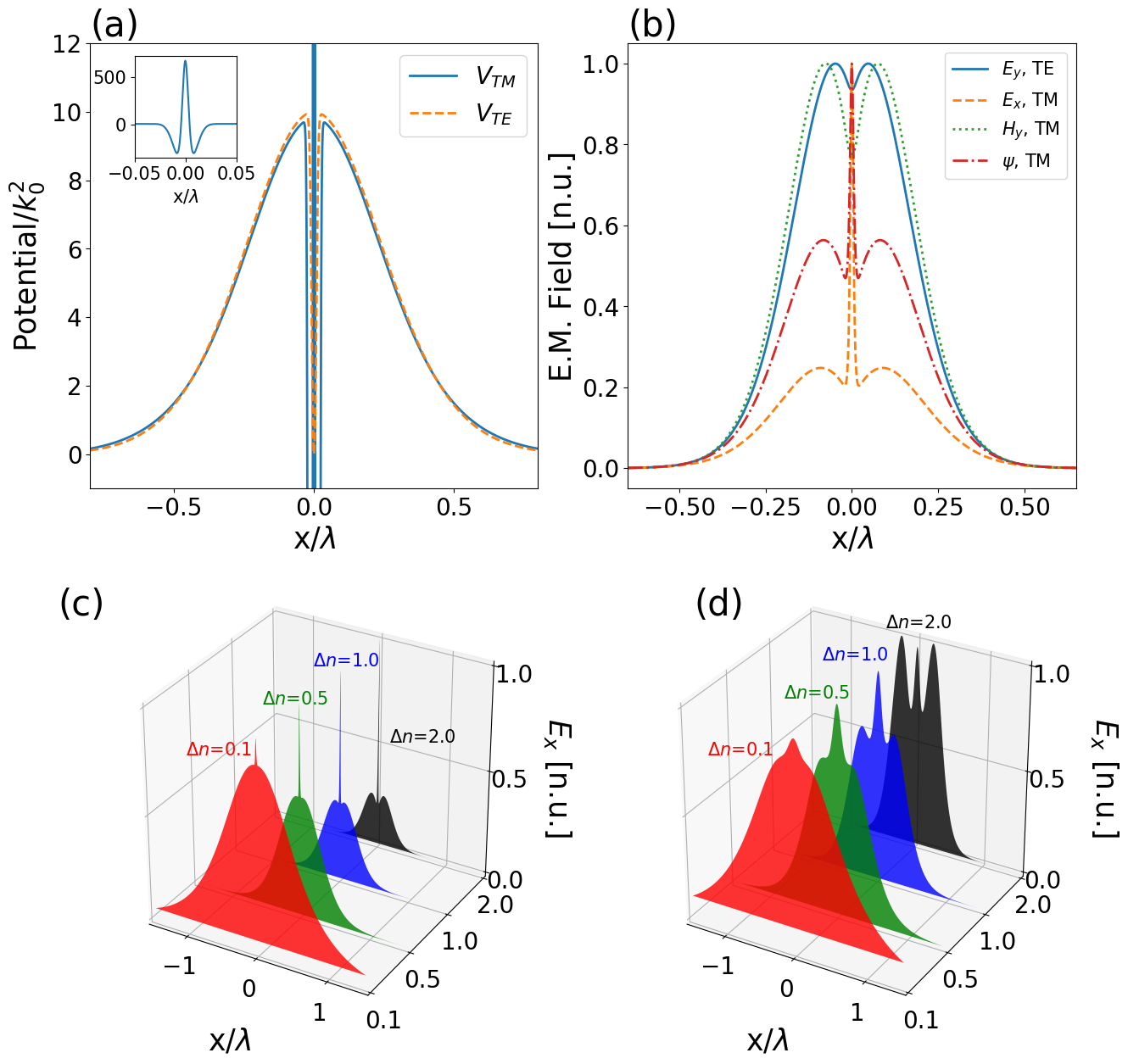}}
\caption{Graded-index slot waveguide. (a) Photonic potential $V/k_0^2-n_0^2$ and (b) electromagnetic field (peak normalized to unity) versus the transverse coordinate $x/\lambda$ for the two polarizations when $\Delta n_0=2$ (corresponding e.g. to Si-on-insulator waveguides). Inset in panel (a): magnification of $V_{TM}$ around the symmetry axis $x=0$. (c-d) Normalized transverse electric field $E_x$ for the TM case versus the maximum refractive index, for a defect of width (c) $w_d/\lambda$=0.01 and (d) $w_d/\lambda=$0.1. All the results are computed for $n_0=1.5$ and $w/\lambda=0.4$.}
\label{fig:GRIN_slot}
\end{figure} \\
\indent Equation~(\ref{eq:potential_TM}) can also be used to design a refractive index distribution $n(x)$ to provide a desired photonic potential $V_{TM}(x)$, let us call it $V_{design}$. For example, $V_{TM}(x)$ can be designed to minimize the bend losses related with the evanescent tails. Then, \eqref{eq:potential_TM} turns into a nonlinear boundary value problem for the profile $n(x)$, where $V_{design}$ plays the role of a forcing term. Standard techniques, such as relaxation algorithms and shooting method, can then be applied to find the solution. In Fig.~\ref{fig:designed_VTM}, we show the results when the target is a Gaussian potential, i.e., $V_{design}=V_0 \exp{\left(-x^2/w^2\right)}+k_0^2n_0^2$. The corresponding refractive index, computed via a standard shooting method with initial conditions $n(x=0)=n_0$ and $\left.\frac{dn}{dx}\right|_{x=0}=0$, is plotted in Fig.~\ref{fig:designed_VTM}(a). The solution is periodic with a sub-wavelength period, modulated around $x=0$ to provide the guiding effect. The obtained profile for $V_{TM}$ [blue solid line in Fig.~\ref{fig:designed_VTM}(b)] is very close to $V_{design}$, except for periodic differences of amplitude $<0.1\%$ [inset in Fig.~\ref{fig:designed_VTM}(b)]. The corresponding fundamental mode both for the TE and the TM polarization and the effective index are plotted in Fig.~\ref{fig:designed_VTM} panel (c) and (d), respectively. For $V_0=2$ the overall variation in $n(x)$ is circa 1.5, in turn yielding a form birefringence $n_{TE}-n_{TM}\approx 0.15$, to be compared with the value of 0.05 obtained for $\Delta n_0=0.5$ for a TE Gaussian nanoguide (see Fig.~\ref{fig:form_birefringence}). FDTD simulations (Fig.~\ref{fig:designed_VTM_fdtd}) confirm that the confinement occurs for both the polarizations, the field profile matching the theoretical predictions. The TE wave undergoes larger coupling losses than the TM polarization, see the solid lines in Fig.~\ref{fig:designed_VTM_fdtd}. 
\begin{figure}[h]
\centering
\fbox{\includegraphics[width=\linewidth]{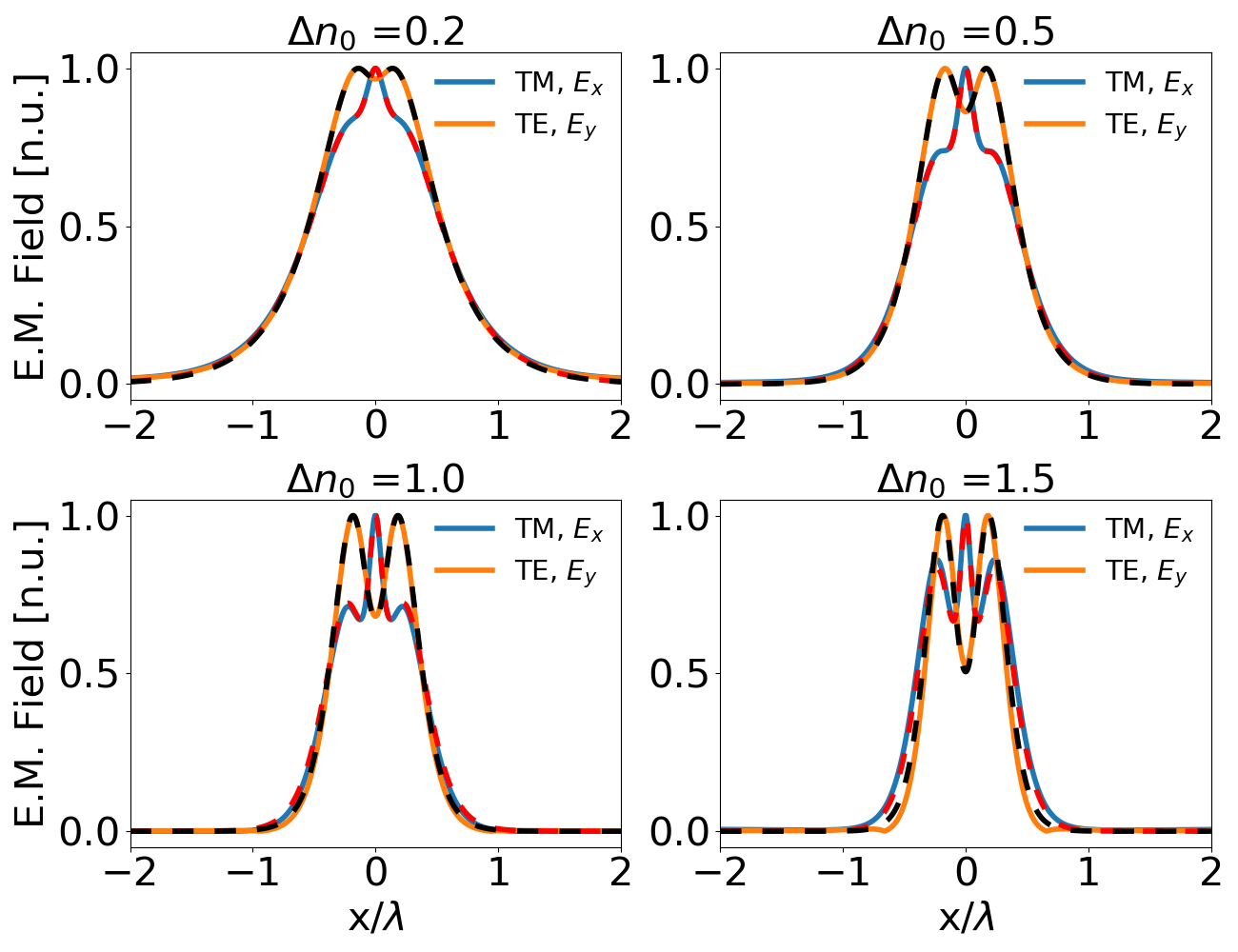}}
\caption{Graded-index slot waveguide. The different panels show the comparison between  the electromagnetic field (peak normalized to unity) obtained analytically (dashed lines) and the FDTD result at 45$\lambda$ (solid lines) for $\Delta n_0$ as marked. Here $n_0=1.5$, $w_d/\lambda=0.1$ and $w/\lambda=0.4$.}
\label{fig:GRIN_slotcomparison}
\end{figure}\\
\indent We now use \eqref{eq:potential_TM} to investigate the light behavior in the presence of a slot waveguide \cite{Almeida:2004,Xu:2004} encompassing a graded-index (GRIN) profile. For the refractive index distribution we make the ansatz $n=n_0+\Delta n_0 e^{-{x^2}/{w^2}} -\Delta n_0 e^{-{x^2}/{w_d^2}}$, that is, a Gaussian waveguide (as the one used in Figs.~\ref{fig:bell_shaped}-\ref{fig:form_birefringence}) with a Gaussian-shaped dip in the center. Thus, we implicitly set $n(x=0)=n_0$. The corresponding photonic potentials for TE and TM polarizations are shown in Fig.~\ref{fig:GRIN_slot}(a). The largest differences between TE and TM modes arise around the central hole, due to the significant gradient in the refractive index. Around $x=0$, the potential for the TM component is dominated by the hole contribution, thus it is fully analogous to Fig.~\ref{fig:bell_shaped}, but inverted in sign. The effect of the central spikes in $V_{TM}$ on the eigenmode can be ascertained by comparing $E_y$ [TE case, solid blue line in Fig.~\ref{fig:GRIN_slot}(b)] with $\psi$ [TM case, dash-dotted red line in Fig.~\ref{fig:GRIN_slot}(b)]. The solution for $\psi$ shows a sharp sub-wavelength peak of width $\approx\lambda/50$ around $x=0$. Recalling that $\text{Re}(\bm{S}\cdot \hat{z})=|\psi|^2$, the TM mode supports a strong local amplification of the carried energy inside the low refractive index core, an important property for optical tweezers \cite{Yang:2009,Descheemaeker:2017}, for example. On the other hand, the magnetic field $H_y=n(x)\psi$ features a dip in $x=0$ [dotted green curve in Fig.~\ref{fig:GRIN_slot}(b)], the latter being deeper than for the TE mode [solid blue line in Fig.~\ref{fig:GRIN_slot}(b)]. Finally, the shape of the transverse electric field in the TM case $E_x=H_y/n^2(x)$ [dashed orange line in Fig.~\ref{fig:GRIN_slot}(b)] is similar to $\psi$, but the prominence of the peak is even stronger than for $\psi$. Noteworthy, the electric field is the quantity to maximize when light-matter interaction needs to be enhanced (supposing a dipolar electric interaction) \cite{Koos:2009}. The prominence of the electric field spike depends strongly on the lateral extension of the central defect, i.e., $w_d$ in our case. Narrower defects [e.g. $w_d/\lambda=0.01$ in Fig.~\ref{fig:GRIN_slot}(c)] yield more prominent peaks [compare with Fig.~\ref{fig:GRIN_slot}(d) where $w_d/\lambda=0.1$] owing to the deepest potential $V_{TM}$ (directly determining $\psi$) and the largest jump in the refractive index (through the relationship between $\psi$ and $E_x$). We verified our predictions simulating the light behavior in the slot waveguide by means of FDTD simulations, the comparison being plotted in Fig.~\ref{fig:GRIN_slotcomparison}. For any value of $\Delta n_0$, the electric fields in the TE and TM case encompass an opposite trend: for TM polarizations the electric field is larger in the low-index core than in the larger index adjacent regions, whereas a dip is observed for TE waves.
%slightly affect the eigenmode  due to their high spatial localization, see Fig.~\ref{fig:GRIN_slot}(b). Conversely, the effect on 

In conclusion, we used the effective photonic potential for TM waves as a new method to design and analyze nanometric optical waveguides, fully accounting for the intrinsic spin-orbit interaction in the sub-wavelength regime. We applied our findings to the design of waveplates and slot waveguides. Our method finds direct application to the investigation of bend losses via transformation optics \cite{Jahani:2018}. Future generalizations include nonlinear effects \cite{Koos:2009} and the extension to the 3D case. Possible implementation in effective medium theories for nano-patterned metamaterials, including nanogratings \cite{Zimmermann:2016,Dai:2016} and dielectric metasurfaces \cite{Chen:2018}, can be envisaged as well.

\textbf{Funding.} Deutsche Forschungsgemeinschaft (DFG) (GRK 2101).

% Bibliography
\bibliography{references}

% Full bibliography added automatically for Optics Letters submissions; the following line will simply be ignored if submitting to other journals.
% Note that this extra page will not count against page length
\bibliographyfullrefs{references}

\end{document}